\documentclass[aps,prl,twocolumn,showpacs,preprintnumbers,runinaddress]{revtex4}
\usepackage{graphicx}
\usepackage{dcolumn}
\usepackage{bm}

\begin{document}

\title{High Efficient Secret Key Distillation for Long Distance Continuous
Variable Quantum Key Distribution}
\author{Yi-Bo Zhao}
\email{zhaoyibo@mail.ustc.edu.cn}
\affiliation{Key Lab of Quantum
Information, University of Science and Technology of China, (CAS),
Hefei, Anhui 230026, China}
\author{Zheng-Fu Han}
\email{zfhan@ustc.edu.cn}
\affiliation{Key Lab of Quantum Information, University of Science and Technology of
China, (CAS), Hefei, Anhui 230026, China}
\author{Jin-Jian Chen}
\affiliation{Key Lab of Quantum Information, University of Science and Technology of
China, (CAS), Hefei, Anhui 230026, China}
\author{You-Zhen Gui}
\affiliation{Key Lab of Quantum Information, University of Science and Technology of
China, (CAS), Hefei, Anhui 230026, China}
\author{Guang-Can Guo}
\affiliation{Key Lab of Quantum Information, University of Science and Technology of
China, (CAS), Hefei, Anhui 230026, China}

\begin{abstract}
The continuous variable quantum key distribution is expected to
provide high secret key rate without single photon source and
detector, but the lack of the secure and effective key distillation
method makes it unpractical. Here, we present a secure
single-bit-reverse-reconciliation protocol combined with secret
information concentration and post-selection, which can distill the
secret key with high efficiency and low computational complexity.
The simulation results show that this protocol can provide high
secret key rate even when the transmission fiber is longer than
150km, which may make the continuous variable scheme to outvie the
single photon one.
\end{abstract}

\maketitle

Recently, many approaches for quantum key distribution have been put
forward, which may be divided into two major types, single photon one \cite%
{1,2} and continuous variable one \cite{3,4,5,6}. The former needs single
photon detector and single photon source or weak coherent source \cite{2},
therefore the high secret key rate becomes a hot topic. Fortunately, the
continuous variable quantum key distribution (CVQKD) is potentially to
provide high secret key rate even for long transmission distance \cite{6},
and only requires the coherent light source as well as the homodyne
detection. Many experiments show the physical possibility of this scheme
\cite{6,7,8}. However, unlike the single photon one, after the quantum
transmission process, the CVQKD only provides continuously distributed raw
key elements \cite{6} that should be converted into binary keys. Although
its potential has been demonstrated, how to distill the secret key is still
an open problem. Firstly, some information will be certainly lost or leaked
during the distillation. Secondly, if taking the concrete distillation into
account, Eve's attack may be much more sophisticated than that considered in
Refs. \cite{3,4,5,6}. When the channel loss is high, the maximum mutual
information between Alice and Bob is only slightly higher than that between
Eve and Bob, and it is possible that even if the CVQKD could be proved
unconditionally secure, any secure binary key could not be distilled.

The only existing reconciliation protocol was proposed by Assche \textit{et
al} \cite{9}, and used in current experimental demonstrations \cite{6}. In
the protocol, they subtly combined the quantization with error correction
and tried the best to reduce publicly exchanged information. However, it is
effective only when the quantum channel loss is small \cite{3,6}. In Ref.
\cite{6}, the error correction block size for the reconciliation was
selected as 50 000, which is already very complex to implement, but only
within 3.1 dB loss, could a secret key be distilled. In fact, the basic
assumption of the previous security proofs of CVQKD is perfect, i.e. an
ideal reconciliation that can convert the continuous-element into common
bits without any auxiliary tactics. However, such reconciliation cannot be
realized under the high loss condition because the error correction it
demands will be too complex to implement \cite{10}. Therefore, a feasible
reconciliation is necessary to show the practical security of the CVQKD.

In the following, we present a single-bit-reverse-reconciliation
protocol to convert the element into common bits, after which large
portion of secret information can still be kept secret from Eve. We
prove that Eve's attack is based on a Markov chain under the
condition of no excess noise, and under the noisy condition, Eve's
tapped additional information can be eliminated. Then we show this
protocol to be secure, which is also a practical security proof for
the CVQKD. In this protocol we concentrate the secret information
into certain bits, expose other useless information and flexibly
employ some tactics, such as post-selection, to keep Eve at a
disadvantage, and then the computational complexity is largely
reduced. With this protocol, 18\% of the theoretical maximum keys
can be distilled at 100km transmission distance (0.01dB loss) and
even at 150km transmission distance (0.001dB loss) the secret keys
can still be distilled.

Here, we discuss the CVQKD scheme of reference \cite{6}, namely the
Gaussian modulated coherent state scheme. Suppose Alice and Bob's
$x$ and $p$ quadratures are $A_{x}$, $A_{p}$, $B_{x}$and $B_{p}$\
respectively, Alice's modulation variance is $V_{A}N_{0}$, where
$N_{0}$ denotes the shot noise variance, and Alice, Bob and Eve's
raw key elements are $a$, $b$ and $c$ respectively. In this scheme
we require that Alice and Bob estimate the conditional entropy and
distribution between them. Then the
single-bit-reverse-reconciliation protocol runs as follows.

\begin{figure}[tbp]
\includegraphics[width=8cm,height=4.5cm]{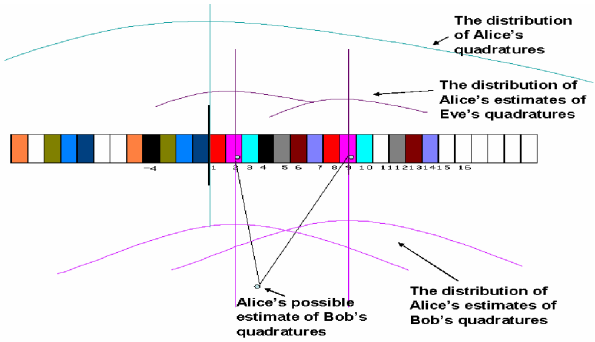}
\caption{An example of the pairing method, where identical colors have been
used to denote the two parties of one pair.}
\label{fig1}
\end{figure}

1. Alice slices her raw key element space into regions, and denotes those
regions with natural numbers in the order as shown in Fig.\ref{fig1}. The
width of each region must be much smaller than the variance of Eve's and
Bob's estimates of Alice's quadratures.

2, Alice sorts her regions into pairs, making sure that each region belongs
to only one pair. The distance between two parties of one pair is $\delta a$%
. For example, we can select pairs by combining 4 with -4, 1 with 8, 2 with
9, \ldots , m with m+7, and so on.

3. Alice converts her key elements into binary bits according to whether
they belong to the left or the right regions of their pairs. If a certain
element belongs to the left region of one pair, Alice designates it as 0,
and if to the right hand side, as 1. Then she tells Bob which pair each of
her raw key elements belongs to, but without exposing her designated bit
results. For example, if one of Alice's key elements belongs to region 2 or
9, then she sends the pairing sign 2-9 to Bob, and assigns it as 0 or 1
correspondingly. The slicing, sorting and assignment methods are publicly
known.

4. According to Alice's pairing sign and his own results, Bob calculates the
conditional probabilities $P(0|b)$, $P(1|b)$ \cite{14} and the bit error
rate (BER) of each of his key elements, where $P(1|b)$ and $P(0|b)$ denote
the probability that Alice assigns her data as 1or 0, while Bob's
measurement is $b$. The BER actually is $\min [P(0|b),P(1|b)]$. Then Bob
divides his elements into 0 or 1 according to which of $P(0|b)$and $P(1|b)$
is bigger. What should be emphasized is that the channel may be binary
asymmetric.

5. After many such communications, Alice and Bob share a series of binary
strings. Then Bob discards those bits with high BER and sorts remained bits
into groups according to the BERs and pairing information to make it
convenient for further error correction, privacy amplification and so on.
For example, he sorts those bits of one pair with approximately identical
BERs into one group. Later, Bob tells Alice which group each of her bits
belongs to.

6. Bob sends the appropriate error correction information of each group to
Alice. After error correction, Alice and Bob share a series of common binary
strings. Then through privacy amplification \cite{11}, they thus finally
have a common secret key.

In the coherent state scheme, depending on Alice's quadratures and Eve's
attack, the entropy of Bob's measurement satisfies \cite{3}:\ \negthinspace $%
H(B_{x}|E_{x})+H(B_{p}|A_{p})\geq 2H_{0},$ where $E_{x}$ represents the
tapped result of Eve, $H(\cdot |\cdot )$denotes the conditional entropy and $%
H_{0}$ is the entropy of a quadrature of the vacuum state. Because Bob
randomly chooses $x$\ or $p$ for his measurement, the secret key rate
guaranteed by the reverse reconciliation is\negthinspace\ $\triangle
I=I(A:B)-I(B:E)=H(B|E)-H(B|A)$, where $I(\cdot |\cdot )$\ denotes the mutual
information. In the same way, we also have $%
H(B_{x}|E_{x},A_{x})+H(B_{p}|A_{p})\geq 2H_{0}.$

We will first discuss the case without excess noise \cite{6}. The states
that Alice sends to Bob are coherent states, so that$\
H(B_{x}|A_{x})+H(B_{p}|A_{p})=2H_{0}$. In the channel estimation, Alice and
Bob should estimate the conditional entropy exactly, which means that any
eavesdropping should maintain\thinspace $H(B|c,A)=H(B|A)$. Then we
obtain:\negthinspace
\begin{equation}
\triangle I=I(A:B|E)  \label{e5}
\end{equation}

From $\triangle I=I(A:B)-I(B:E)$ and Eq.(\ref{e5}) we have:\negthinspace
\begin{equation}
H(B,E|A)=H(E|A)+H(B|A)  \label{e6}
\end{equation}

On the other hand, Alice sends Bob the coherent state $|a\rangle $, which is
attenuated to $|\sqrt{G}a\rangle $ by the time it arrives at Bob when the
channel transmission is $G$. Thereby, under the case of Gaussian modulation,
the distributions of $a$ and $b$ can be written as:

\begin{equation}
P(a)=\frac{1}{\sqrt{2\pi V_{A}N_{0}}}\exp (-\frac{a^{2}}{2V_{A}N_{0}})
\label{e7}
\end{equation}

\negthinspace \negthinspace \negthinspace
\begin{equation}
\!P(b|a)=\frac{1}{\sqrt{2\pi N_{0}}}\exp [-\frac{(\sqrt{G}a-b)^{2}}{2N_{0}}]
\label{e8}
\end{equation}

The coherent state guarantees the conditional distribution $P(b|a)$ to be
the distribution with the minimum shot noise, and since Eve cannot know
which quadrature Bob will choose to measure, after many communications Alice
and Bob can employ, for example, the normality test to limit Eve's tapped
information. Any of her attacks should maintain the conditional distribution
of Eq.(\ref{e8}), which means $P(b|a)=P(b|a,c)$.

We can then obtain $P(b,c|a)=P(c|a)P(b|a,c)=P(c|a)P(b|a)$\ and $%
P(c,a|b)=P(c|a)P(b|a)P(a)/P(b)=P(c|a)P(a|b)$.

Thus\negthinspace
\begin{equation}
P(c|b)=\int P(c|a)P(a|b)da  \label{e9}
\end{equation}%
{}\negthinspace \negthinspace \negthinspace \negthinspace \negthinspace
\negthinspace
\begin{equation}
P(b|c)=\int P(b|a)P(a|c)da  \label{e10}
\end{equation}%
Eqs.(\ref{e5}) (\ref{e6}) (\ref{e9}) and (\ref{e10}) show that under the
condition that Alice's quadratures are known, Bob's and Eve's measurements
are independent. They also illustrate that, for reverse CVQKD, if there is
no excess noise, any eavesdropping of Bob's information is equivalent to a
Markov chain of $Eve\rightarrow Alice\rightarrow Bob$, which means that for
the reverse scheme, the information can be regarded as having been sent by
Bob, received by Alice, and then transmitted to Eve by Alice. All of Eve's
information about Bob is obtained from Alice. To estimate Bob's result, at
first Eve should estimate Alice's quadrature, which is restricted by the
direct reconciliation schemes \cite{12}

After Alice announces her paring information, Eve can only know that Alice's
sent state is equivalent to $|a_{0}\rangle \ $or $|a_{0}+\delta a\rangle $,
where $|a_{0}\rangle $ describes Alice's certain sent state. After Bob
publics his grouping information, Eve knows the channel properties between
Alice and Bob, but her estimates of Bob's result are still based on her
estimates of Alice's quadrature. To Eve, the probability of Bob's assignment
results is\negthinspace\ $P(b\rightarrow 1|c)=P(|a_{0}+\delta a\rangle
|c)P(b\rightarrow 1||a_{0}+\delta a\rangle )+P(|a_{0}\rangle
|c)P(b\rightarrow 1||a_{0}\rangle )$, where $b\rightarrow 1$ correspond to
Bob assigning his data as 1 and $P(|a_{0}\rangle |c)$ describes the
probability that Alice has sent $|a_{0}\rangle $ while Eve's result is $c$. $%
\min [P(b\rightarrow 1|c),P(b\rightarrow 0|c)]$ actually denotes Eve's BER.
It shows that after the assignment the binary channel is a binary Markov
chain of $Bob\rightarrow Alice\rightarrow Eve$. If Eve cannot estimate
Alice's result correctly, she cannot estimate Bob's correctly. The final
amount of practical secret keys can be given by%
\begin{equation}
\triangle I=n\sum_{G}P(G)[I_{BG}(A:B)-I_{BG}(E:B)]  \label{e12}
\end{equation}%
where $n$ denotes the number of raw key elements, $\sum_{G}$ means summation
over the groups, $P(G)$ describes the probability of the groups, and $%
I_{BG}(\cdot :\cdot )$ denotes the mutual information of a group after their
exposure. We see that because of the Markov chain $I_{BG}(A:B)\geq
I_{BG}(E:B)$ is always maintained \cite{13}. Through Alice's pairing
information and Bob's grouping information, the transmission characteristics
of the binary channel between Alice and Bob can be given. The properties of
the channel between Eve and Alice are discussed in many direct
reconciliation schemes \cite{12}. Then the maximum information tapped by Eve
can also be given.

If Bob sorts those bits of a pair with roughly the same BER into one group,
for a certain group and pair, $I_{BG}(A:B)=H_{BG}(A)-\sum_{Group}P_{G}(b)%
\Gamma \lbrack P(0|b)]$, where $H_{BG}(\cdot )$ denotes the Shannon entropy
of the distribution of 0 and 1 in a certain group, $\sum_{Group}$ denotes
summation in the group, $P_{G}(\cdot )$ is the distribution within the
group, depending on Bob's grouping method, and $\Gamma (\lambda )=-\lambda
\log _{2}\lambda -(1-\lambda )\log _{2}(1-\lambda )$ is the Shannon entropy.

The mutual information of a group between Eve and Bob can be given by $%
I_{BG}(E:B)=H_{BG}(B)-\int P_{G}(c)\Gamma \lbrack P(b\rightarrow 1|c)]dc$.
The minimum noise on Alice's side in estimating Eve's quadrature is given by
reference \cite{12}: $N_{E}=N_{0}+\frac{GN_{0}}{1-G}=\frac{N_{0}}{1-G}$. For
the minimum Gaussian noise case, we have the conditional distributions: $%
P(c|a)=\frac{1}{\sqrt{2\pi \alpha N_{E}}}\exp [-\frac{(\sqrt{\alpha }a-c)^{2}%
}{2\alpha N_{E}}],$ where $\alpha $ is a parameter which do not affect the
final results. Then by the grouping information, $P_{G}(c),$ $%
P(|a_{0}\rangle |c)$ and consequently $I_{BG}(E:B)$ can be obtained. The
concrete calculation is complex and depending on Bob's grouping method.
Here, we only give the simulation results.

\begin{figure}[tbp]
\includegraphics[width=8.0cm,height=5.0cm]{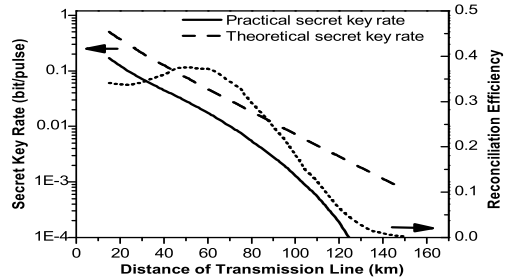}
\caption{Simulation results for secret key rate versus transmission
line distance. Dashed line: theoretical key rate from Ref.
\protect\cite{6}; solid line: practical key rate from our simulation
result using the above method; dotted line: efficiency defined as
the ratio of the practical to the theoretical key rates}
\label{fig2}
\end{figure}

The simulation results of the binary secret key rate as a function of the
channel transmission distance are shown in Fig.\ref{fig2}, where we have set
$\delta a=(1+0.02L)\sqrt{N_{E}}$, $V_{A}=500$ and the width of the region $%
\delta a/7$ and $L$ is the transmission distance in the unit of km. The
pairing method is that shown in Fig.\ref{fig1}, and the bits of both Alice
and Bob are sorted into groups with roughly the same BER, while those bits
with too high a BER, higher than 40\%, are discarded.
In the simulation we see that the maximum efficiency, defined as the
ratio of the practical to the theoretical secret key rates, is about
37\%. At the distance of 100km, the efficiency is 18\%, the BER
between Alice and Eve is higher than 4\% and the secret information
per final bit carries is approximately equal that per key element
carries. Even for long distances approaching 150km the efficiency is
about 0.15\%. While in Ref. \cite{6}, they distilled the key only
within 3.1dB loss (15km fiber). For long distances the BER between
Alice and Bob is increased, so many bits should be discarded to
reduce the computational complexity, at the cost of lower
efficiency.

When the channel has excess noise $\xi N_{0}$, we have $V_{B|A}=(1+G\xi
)N_{0}$ \cite{6}, where $V_{\cdot |\cdot }$ denotes the conditional
variance, so $H(B|A)=H_{0}+0.5\log _{2}(1+G\xi )=H_{0}+H^{\prime }$. Under
this condition through entropy estimation Alice and Bob may not be able to
find a lurking Eve when $H(B|A,c)\geq H(B|A)-2H^{\prime }$, and then the
secret key rate becomes $\triangle I=I(A:B|E)-2H^{\prime }$. Obviously, this
case is not always a Markov chain. In this situation Eve can do entangling
cloner attack, using a noiseless line to replace the noise one between Alice
and Bob and injecting an entangled beam into the line to tap more of Bob's
information \cite{6}. However, under this case, the maximum variance change
caused by the entangled beam cannot exceed that of the excess noise, so the
maximum additional information Eve can attack is $I_{add}=0.5\log
_{2}[V_{B|A}/\min (V_{vacuum})]=0.5\log
_{2}(V_{B|A}^{2}/N_{0}^{2})=2H^{\prime }$ bits \cite{13,15}, where $\min
(V_{vacuum})$ denotes the minimum variance of vacuum noise. Moreover,
previous analyses show that $I(A:B|E)$ is only determined by the Markov
chain. Thus, this process is equivalent to that besides through the previous
Markov chain Eve can obtain additional information of $2H^{\prime }$ bits
from Bob. It can be proved that the exactly same result in reference \cite{6}
can be derived by this way. Therefore, at first Alice and Bob can suppose
the channel is still a Markov chain, and estimate the maximum information
tapped by Eve. If the results satisfy $\triangle I>2H^{\prime }$, Alice and
Bob can still distill a secret key, but the final amount should be
subtracted by $2nH^{\prime }$ bits. Then the amount of secret keys becomes: $%
\triangle I=n\{{\sum_{G}P(G)[I_{BG}(A:B)-I_{BG}(E:B)]-2H^{\prime }\}}$.

What should be noticed is that if there is excess noise, the conditional
distribution of Bob and Eve's measurements will all be changed
correspondingly. The $N_{E}$ will become $N_{0}+N_{0}/(\frac{1-G}{G}+\xi )$
\cite{12}. If the efficiency is 100\%, only when $\xi <0.5$, can the secret
key be distilled \cite{6}. If the reconciliation efficiency is $\gamma $, it
requires $\xi <0.5\gamma $, which can be improved by choose proper
post-selection.

In this protocol, after Alice and Bob's exposure, the assignment process is
equivalent to binary modulation, but Eve's attack is still limited by CVQKD.
To reduce the computational complexity, we should concentrate the secret
information into certain bits and increase the difference between Eve's
minimum BER and Alice's BER \cite{10}. In this protocol we see that if there
is no excess noise the channel is a Markov chain, so if the BER between
Alice and Eve is maintained nonzero, then the BER between Eve and Bob is
certainly larger than that between Alice and Bob \cite{16}. Alice and Bob
can use post-selection, discarding those bits with a high BER between them.
Then the BER of remained bits between Eve and Bob will be sufficiently
larger than that between Alice and Bob, so the complexity of the error
correction can be significantly reduced \cite{10}. For example, if the BER
between Alice and Bob of remained bits is below 15\%, while that between Eve
and Alice is maintained to be 25\%, then the difference between the BER
between Eve and Bob and that between Alice and Bob is larger than 17\%. In
addition, to reduce the computational complexity and increase the secret key
rate we should choose an appropriate value for $\delta a$, since $\delta a$
affects both the BER between Alice and Bob and that between Alice and Eve.
We slice Alice's raw key element space into very small regions, mainly
because by doing so Alice and Bob can estimate their BERs exactly. Moreover,
Bob should sort his binary strings into different groups according to the
BERs, because the bits with different BERs may contribute different amounts
of secret key, and require different error corrections.

In this letter, we show that under the condition of no excess noise
Eve's attack is based on a Markov chain in the reverse
reconciliation CVQKD scheme, and the excess noise will only cause
additional information loss, which can be easily subtracted in the
final result. Then we propose a secure
single-bit-reverse-reconciliation protocol, with low computational
complexity. The numerical simulation results show that almost 18\%
of the theoretical maximum secret information can be distilled by
the protocol with low computational complexity even when the
transmission fiber is longer than 100km.

\textbf{Acknowledgement}: Special thanks are given to Professor Ling-An Wu
and Professor P. Grangier. We also thank F. Zhang for valuable discussions
on classical quantization, and bit assignment. Thanks are also due to X. N.
Ji, Q. Fu, B. Zhu, J. Chen, H. Wen and Y. Liu for their help in the
simulations. Support from Science Foundation of China under Grant No.
60537020 and No. 60121503 and the Knowledge Innovation Project of Chinese
Academy of Sciences.


\begin{thebibliography}{99}
\bibitem{1} C. H. Bennett and G. Brassard. \textit{Proceedings of IEEE
International Conference on Computers, Systems, and Signal Processing (IEEE,
New York, 1984)}, pp. 175-179.

\bibitem{2} W. Y. Hwang, Phys. Rev. Lett. \textbf{91}, 057901 (2003).

\bibitem{3} F. Grosshans, N. J. Cerf, Phys. Rev. Lett. \textbf{92}, 047905
(2004).

\bibitem{4} S. Iblisdir, G. VanAssche and N. J. Cerf, Phys. Rev. Lett.
\textbf{93}, 170502 (2004).

\bibitem{5} C. Weedbrook, A. M. Lance, W. P. Bowen, T. Symul, T. C. Ralph
and P. K. Lam, Phys. Rev. Lett. \textbf{93}, 170504 (2004).

\bibitem{6} F. Grosshans, G. VanAssche, J. Wenger, R. Brourl, N. J. Cerf and
P. Grangier, Nature \textbf{421}, 238 (2003).

\bibitem{7} M. Legre, H. Zbinden and N. Gisin, arXiv: quant-ph/0511113
(2005).

\bibitem{8} J. Lodewyck, T. Debuisschert, R. Tualle-Brouri and P. Grangier,
Phys. Rev. A \textbf{72}, 050303(R) (2005).

\bibitem{9} G. VanAssche, J. Cardinal and N. J. Cerf, IEEE Trans. Inform.
Theory, \textbf{50}, 394 (2004).

\bibitem{10} Y. B. Zhao, Y. Z. Gui, J. J. Chen, Z. F. Han, G. C. Guo, arXiv:
quant-ph/0602019 (2006).

\bibitem{11} C. H. Bennett, G. Brassard, C. Crepeau, U. M. Maurer, IEEE
Trans. Inform. Theory, \textbf{41}, 1915 (1995).

\bibitem{12} F. Grosshans, P. Grangier, Phys. Rev. Lett. \textbf{88}, 057902
(2002).

\bibitem{13} C. E. Shannon, Bell Syst. Tech. J. \textbf{27}, 379 (1948).

\bibitem{14} $P(1|b)=1/\{1+\exp \{\frac{\delta a}{2V_{A}N_{0}}%
[(1+GV_{A})(2a_{0}+\delta a)-2\sqrt{G}V_{A}b]\}\}$.

\bibitem{15} $V_{B|A}\min (V_{vacuum})=N_{0}^{2}$.

\bibitem{16} For symmetry binary channals, $%
e_{EB}=e_{EA}+e_{AB}-2e_{EA}e_{AB}$, where $e_{\cdot \cdot }$ is the BER.
\end{thebibliography}
\end{document}